# Distributed Multi-task Formation Control under Parametric Communication Uncertainties

Dongkun Han and Dimitra Panagou

*Abstract*— Formation control is a key problem in the coordination of multiple agents. It arises new challenges to traditional formation control strategy when the communication among agents is affected by uncertainties. This paper considers the robust multi-task formation control problem of multiple non-point agents whose communications are disturbed by uncertain parameters. The control objectives include 1. achieving the desired configuration; 2. avoiding collisions; 3. preserving the connectedness of uncertain topology. To achieve these objectives, firstly, a condition of Linear Matrix Inequalities (LMIs) is proposed for checking the connectedness of an uncertain communication topology. Then, by preserving the initial topological connectedness, a gradient-based distributed controller is designed via Lyapunov-like barrier functions. Two numerical examples illustrate the effectiveness of the proposed method.

## I. INTRODUCTION

Multi-agent systems consist of a group of intelligent agents or subsystems that cooperate to achieve collective goals through mutual communications. Due to its broad applications in various areas, a number of cooperative problems have been introduced, such as consensus, flocking, rendezvous and coverage [1]–[4]. Among these problems, the *formation control* aims to achieve a desired configuration specified by relative inter-agent distances. This problem attracts much attention in recent years because of its wide range of practical implementations, e.g., aircrafts coordination, multi-robot path planning, and air-defense of warships (for more applications, please find in [5]).

Due to the great amount of literature in formation control, we focus on whether the communication network of multiple agents is disturbed or not, rather than an exhaustive review from all aspects (excellent surveys of formation control can be found in [6]–[8]). For the case of communication network without disturbances, efficient methods are proposed by using algebraic graph theory, Lyapunov stability theorem, LaSalle invariance principle, behavior-based approaches, etc [9], [10]. However, in real world, the communication among agents is usually disturbed by sorts of uncertainties, under which the methods developed for disturbance-free cases are usually not applicable [11].

In order to improve the robustness to the uncertainties in communication, different uncertain models and useful

The authors are with the Department of Aerospace Engineering, University of Michigan, 1320 Beal Ave, Ann Arbor, MI 48109, USA. E-mail: {dongkunh,dpanagou}@umich.edu.

This work is supported by the Automotive Research Center (ARC) in accordance with Cooperative Agreement W56HZV-14-2-0001 U.S. Army TARDEC in Warren, MI, USA, the NASA Grant NNX16AH81A, and an Early Career Faculty Grant from NASA's Space Technology Research Grants Program.

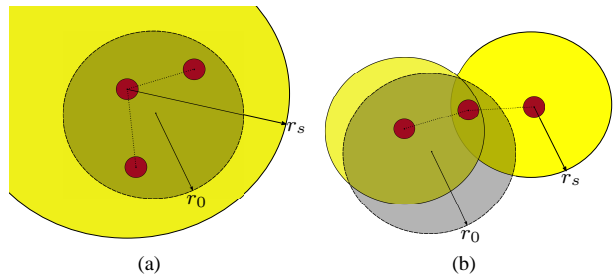

Fig. 1. Different strategies for preserving the connectedness. $r_0$ is the radius of a chosen working place; $r_s$ is the radius of sensing of each agent.

methods have been proposed recently [11]–[16]. According to the types of uncertainties, these models can be roughly divided into two categories: *Additive norm-bounded* uncertainties and *additive stochastic* uncertainties. In the first category, it is assumed that an uncertainty is added to the control input for each agent, where the $l_2$-norm or $l_\infty$-norm of this additive uncertainty is bounded, and Lyapunov stability theorem is employed for controller design [12]–[14]. In the second category, Guassian noises or white noises are added to the measurement or to the control input, reflecting the unreliable information exchange [11], [15]. However, additive uncertainties cannot meet all practical demands, e.g., (as a motivating example for this work) in an aerospace multi-robot network, the communication among robots are disturbed by parameters like temperature, radiation, magnetic field intensity, such that the value of information exchange can be modeled as a polynomial function of these parameters.

Besides additive uncertainty, the existing literature usually assumes that the topology has a spanning tree, or it is connected. Though this condition is essential, the network cannot be always connected if there is no control input guaranteeing the connectedness. Thus, researchers start focusing on the connectedness maintenance in multi-agent networks. One strategy for connectedness maintenance is to constrain all the agents in a selected working place by using Lyapunov-like barrier functions [17], and the network is strongly connected if $r_s \geq 2r_0$ as shown in (a) of Fig. 1. Another strategy is to use a potential function to preserve edges once they appear in the network [10], [18]–[24]. The latter strategy does not require constraining the agents in a work place, hence less conservative than the first one, as shown in Fig. 1. In [22], for uncertainty-free case, a decentralized navigation function is introduced for connectedness maintenance and collision avoidance. In [19], a robust rendezvous problem is studied with unknown dynamics and bounded disturbances, but without considering collision avoidance.

Motivated by aforementioned results, based on our previous work [17], this paper considers a robust formation control problem of uncertain networks, whose communication weights are generic polynomial functions of uncertain parameters. The main contributions of this paper are:

- Rather than calculating the eigenvalues of Laplacian matrix, a necessary and sufficient condition is provided for checking the connectedness of uncertain networks. Based on the real Positivestellensatz, a relaxed condition is given via sum-of-squares programming. Then, by using the square matrix representation, a solvable condition of LMIs is obtained for checking the connectedness of uncertain network (Section III.A).
- Different from [20] and [23], a new Lyapunov-like barrier function is introduced, which guarantees a bounded control input. Then, a gradient-based controller is designed for solving the robust formation problem, with considering both connectedness maintenance and collision avoidance (Section III.B).

## II. PRELIMINARIES

Notations: $\mathbb{N}, \mathbb{R}$: natural and real number sets; $\mathbb{R}_+$: positive real number set; $A^T$: transpose of $A$; $A > 0$ ($A \geq 0$): symmetric positive definite (semidefinite) matrix $A$; $A \otimes B$: Kronecker product of matrices $A$ and $B$; $\text{diag}(a)$: a square diagonal matrix with the elements of vector $a$ on the main diagonal; $\|a\|$: Euclidean norm or $l_2$ norm of vector $a$. $1_n$: $n \times 1$ vector with all the entries equal to 1; $I$: identity matrix (of size defined by the context); $\deg(f)$: degree of polynomial function $f$; $\text{Deg}(M)$: degree of matrix polynomial function $M$, i.e., $\max(\deg(M_{ij}))$; $(*)^T AB$ in a form of Square Matrix Representation: $B^T AB$. Let $\mathcal{P}$ be the set of polynomials and $\mathcal{P}^{n \times m}$ be the set of matrix polynomials with dimension $n \times m$. A polynomial $p(x) \in \mathcal{P}$ is nonnegative if $p(x) \geq 0$ for all $x \in \mathbb{R}^n$. A useful way of establishing $p(x) \geq 0$ consists of checking whether $p(x)$ can be described as a sum of squares of polynomials (SOS), i.e., $p(x) = \sum_{i=1}^{k} p_i(x)^2$ for some $p_1, \ldots, p_k \in \mathcal{P}$. The set of SOS polynomials is denoted by $\mathcal{P}_{\text{SOS}}$. A symmetric matrix polynomial $M(x) \in \mathbb{R}^{l \times l}$ is SOS if there exist matrix polynomials $M_1(x), M_2(x), \ldots$ such that $M(x) = \sum_i M_i(x)^T M_i(x)$. The set of SOS matrix polynomials is denoted by $\mathcal{P}_{\text{SOS}}^{l \times l}$.

### A. Model Formulation

In this paper, we consider a second-order system of multiple agents described by:

$$\begin{array}{rcl} \dot{x}_i(t) & = & \rho_i(t) \\ \dot{\rho}_i(t) & = & u_i(t), \quad i \in \mathcal{N}, \end{array} \quad (1)$$

where $\mathcal{N} = \{1, \ldots, N\}$, $x_i(t) \in \mathbb{R}^n$ denotes the position state, $\rho_i(t) \in \mathbb{R}^n$ denotes the velocity state, and $u_i(t) \in \mathbb{R}^n$ denotes the control input on $i$-th agent. In the sequel, we will omit the arguments $t$ and $x$ of functions whenever possible for the brevity of notations.

Based on algebraic graph theory [3], a network of multiple agent can be described by a *weighted undirected dynamic graph* $\mathcal{G}(t) = (\mathcal{A}, \mathcal{E}(t), G)$ with the set of nodes $\mathcal{A} = \{A_1, \ldots, A_N\}$, the set of undirected edges $\mathcal{E}(t) = \{(A_i, A_j) | A_i, A_j \in \mathcal{A}\}$, and the *weighted adjacency matrix* $G = (G_{ij})_{N \times N}$. Th set of edges $\mathcal{E}(t)$ is constructed similar to [19] as shown in Fig. 2. A graph $\mathcal{G}(t)$ is *connected* at

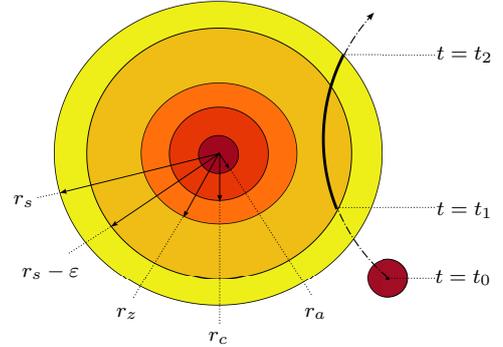

Fig. 2. The model of a single agent and the switching strategy of edges between two agents: $r_a$ is the radius of each agent; $r_c$ is the radius of collision avoidance region; $r_z$ is the radius of region that the control with collision avoidance objective is active; $r_s$ is the radius of sensing range; constant $\epsilon \in [0, r_s - r_z]$ is a distance parameter for the hysteresis in adding new edges. The solid line for $t \in (t_1, t_2)$ depicts the part of trajectory when there is an edge between the two agents.

time $t$ if there is a path between any pair of distinct nodes $A_i$ and $A_j$ in $\mathcal{G}(t)$. The *Laplacian matrix* is given as

$$L(t) = \Delta(t) - G(t) \quad (2)$$

where $\Delta(t) = \text{diag}(\sum_{j=1}^{N} G_{ij}(t))$. A relationship between $L(t)$ and the connectedness of $\mathcal{G}(t)$ is given in [3], [25]:

*Lemma 1:* Let $\lambda_1(L(t)) \leq \lambda_2(L(t)) \leq \cdots \leq \lambda_N(L(t))$ be the ordered eigenvalues of $L(t)$. Then, $1_N$ is an eigenvector of $L(t)$ with the corresponding eigenvalue $\lambda_1(L(t)) = 0$. Moreover, $\lambda_2(L(t)) > 0$ if and only if $\mathcal{G}(t)$ is connected. $\square$

### B. Problem Formulation

In this paper, the parametric uncertainties are considered in communications among agents. Specifically, it is supposed that the weighted adjacency matrix $G$ is affected by uncertain parameters. We denote such a matrix as $G(t, \theta)$ where $\theta \in \mathbb{R}^r$ is an uncertain vector constrained as

$$\theta \in \Omega, \ \Omega = \{\theta \in \mathbb{R}^r : \ s_i(\theta) \geq 0 \ \forall i = 1, \ldots, h\} \quad (3)$$

for some functions $s_1, \ldots, s_h : \mathbb{R}^r \to \mathbb{R}$. In the sequel, we will assume that the entries of $G(t, \theta)$ and $s_1(\theta), \ldots, s_h(\theta)$ are polynomials.

*Remark 1:* The type of uncertain model introduced above is more general compared to the additive norm-bounded uncertainties in [12]–[14]. Besides the motivating example given in Section I, more practical implementations in intelligent transportation and under-water vehicles control can be found in [6]. $\square$

For a decentralized controller of agent $i$, the control law relies on the local information of agent $i$. Specifically,

$$u_i = \sum_{j \in \mathcal{N}_i^s(t)} f\Big(x_i(t) - x_j(t),\ \rho_i(t) - \rho_j(t),\ G_{ij}(t, \theta)\Big) \quad (4)$$

where $\mathcal{N}_i^s(t) = \{j|\ (A_i, A_j) \in \mathcal{E}(t)\}$ is the neighborhood set of agent $i$ (in the sensing range of agent $i$). Different with flocking and rendezvous problems, some useful sets are defined for formation control: $\mathcal{N}_i^f$ is the neighborhood set to agent $i$ in the desired formation configuration, i.e., $\mathcal{N}_i^f = \{j|\ (A_i, A_j) \in \mathcal{E}^f,\ x_i - \tau_i - (x_j - \tau_j) = 0\}$, where $\tau_i$ is the ideal displacement of agent $i$ in the desired formation configuration, whose edge set is $\mathcal{E}^f$; We also define sets $\mathcal{N}_i^{sf}(t) = \{j|\ j \in \mathcal{N}_i^s(t),\ j \in \mathcal{N}_i^f\}$ and $\mathcal{N}_i^{sz}(t) = \{j|\ j \in \mathcal{N}_i^s(t),\ \|x_i - x_j\| < r_z\}$, which will be used in Section III.

Now, let us propose the *robust multi-objective formation control problem* we are concerned with:

*Problem 1:* Under the dynamics (1) and the effect of $\theta$ given in (3), find a controller $u_i$ in the form of (4) such that
1) $\lim_{t\to\infty} \|(x_i(t) - \tau_i) - (x_j(t) - \tau_j)\| = 0$, and $\lim_{t\to\infty} \|\rho_i - \rho_j\| = 0$, for all $i \in \mathcal{N}$ and $j \in \mathcal{N}_i^f$.
2) $\mathcal{G}(t)$ is connected, for all $t > t_0$, where $t_0$ is the initial time.
3) $\|x_i(t) - x_j(t)\| > d_s$, for all $t > t_0$, where $d_s$ denotes a user-defined safety distance for collision avoidance. □

For this problem, we assume that:
- Assumption 1: The desired configuration given by $\tau_i$ is achievable, i.e., $r_z \leq \|\tau_i - \tau_j\| \leq r_s - \varepsilon$, for all $i \in \mathcal{N}$, $j \in \mathcal{N}_i^f$. In other words, the desired distance between agent $i$ and agent $j \in \mathcal{N}_i^f$ is always between $r_s - \varepsilon$ and $r_z$.
- Assumption 2: The neighbor set of agent $i$ at time $t_0$ satisfies $\mathcal{N}_i^f \subseteq \mathcal{N}_i^s(t_0)$, which means that the desired topology is contained in the initial graph.
- Assumption 3: To achieve both objectives of collision avoidance and connectedness maintenance, we require $r_s - \|\tau_{ij}\| > d_s + \|\tau_{ij}\|$, for all $i, j \in \mathcal{N}$.

## III. MAIN RESULTS

In this section, a condition for checking the connectedness of uncertain networks is provided, and a distributed controller is designed to solve Problem 1. The main idea is to preserve the connectedness of initial uncertain graph, and then use a gradient-based controller based on properly choosing barrier Lyapunov functions.

### A. Checking the Connectedness of Uncertain networks

The connectedness of uncertainty-free network can be established by checking whether $\lambda_2(L)$ is positive from Lemma 1. However, it is not easy to calculate $\lambda_2(L(t^*, \theta))$ at a chosen time $t^*$ under the disturbance of $\theta$ given in (3). In this subsection, we propose a method of Linear Matrix Inequalities (LMIs) to check the connectedness of an uncertain graph $\mathcal{G}(t^*)$ at a fixed time $t^*$. We omit the argument $t$ in this subsection.

First, a necessary and sufficient condition is proposed for establishing the connectedness of uncertain networks:

*Lemma 2:* Consider a matrix $M \in \mathbb{R}^{N \times N-1}$ satisfying $\text{img}(M) = \ker(1_N^T)$, then, a new matrix is constructed based on the Laplacian matrix as:

$$\widehat{L}(\theta) = M^T L(\theta) M \quad (5)$$

where $\text{img}(A)$ denotes the image of matrix $A$, $\ker(A)$ denotes the null space of $A$, $1_N$ denotes a $N$ dimensional column vector with all the entries equal to 1. The graph $\mathcal{G}$ is connected under the effect of parametric uncertainties $\theta$ if and only if there exists a symmetric definite matrix function $P(\theta): \mathbb{R}^r \to \mathbb{R}^{N-1 \times N-1}$ such that

$$P(\theta)\widehat{L}(\theta) + \widehat{L}(\theta)^T P(\theta) > 0,\ \forall \theta \in \Omega. \quad (6)$$

*Proof:* From Lemma 1 and (2), one has that $1_N$ is an eigenvector of $L(\theta)$ corresponding to the eigenvalue zero. In addition, observe that $M^T L(\theta) M$ has the same eigenvalues of $L(\theta)$ except that the algebraic multiplicity of the eigenvalue zero has been decreased of one. i.e. $\text{spc}(\widehat{L}(\theta)) \cup \{0\} = \text{spc}(L(\theta))$, where $\text{spc}(A)$ is the spectrum of matrix $A$. Consider the following dynamical system

$$\dot{\hat{x}}(t) = -\widehat{L}(\theta)\hat{x}(t), \quad (7)$$

one has that the linear hull of vector $1_{N-1}$, i.e., $\text{span}(1_{N-1})$ is the equilibrium point of (7). It yields that the condition $0 < \lambda_2(L(\theta)) \leq \lambda_3(L(\theta)) \leq \ldots \lambda_N(L(\theta))$ is equivalent to the statement that (7) is asymptotically stable. Let us consider a parameter-dependent Lyapunov function $V(\hat{x}, \theta) = \hat{x}^T P(\theta)\hat{x}$ with $P(\theta) > 0$. From Lyapunov stability theorem, one has that (7) is asymptotically stable for all $\theta \in \Omega$ if and only if $V(\hat{x}, \theta) > 0$ and $\dot{V}(\hat{x}, \theta) < 0$. Moreover, observe $\dot{V}(\hat{x}, \theta) = -\hat{x}^T(P(\theta)\widehat{L}(\theta) + \widehat{L}(\theta)^T P(\theta))\hat{x} < 0$ is equivalent to the condition of (6), it yields that (6) holds if and only if $L(\theta)$ has exactly one simple eigenvalue 0 and all the other eigenvalues have positive parts, i.e., $\lambda_2(L(\theta)) > 0$ for all $\theta \in \Omega$, which completes this proof. □

Note that the matrix polynomial inequalities (6) is not easy to be established, thus a tractable condition is provided based on the Real Positivestellensatz [26] as follows:

*Lemma 3:* Condition (6) holds if there exist matrix polynomials $P(\theta)$, $R_i(\theta)$, for $i = 1, \ldots, h$, and positive scalars $c_1 > 0$ and $c_2 > 0$ such that

$$\begin{cases} R_i(\theta) \in \mathcal{P}_{\text{SOS}}^{N-1 \times N-1} \\ P(\theta) - c_1 I_{N-1} \in \mathcal{P}_{\text{SOS}}^{N-1 \times N-1} \\ H(\theta) - c_2 I_{N-1} \in \mathcal{P}_{\text{SOS}}^{N-1 \times N-1} \end{cases} \quad (8)$$

where $\mathcal{P}_{\text{SOS}}^{N-1 \times N-1}$ is the set of SOS matrix polynomials introduced in Section II,

$$H(\theta) = P(\theta)\widehat{L}(\theta) + \widehat{L}(\theta)^T P(\theta) - \sum_{i=1}^{h} R_i(\theta) s_i(\theta), \quad (9)$$

and $\theta$, $s_i$ are defined in (3).

*Proof:* Assume that (8) holds, one has $P(\theta) > 0$ and $H(\theta) > 0$. In addition, $R_i(\theta)$ are SOS matrix polynomials,

and $s_i(\theta) > 0$ from (3). Based on the Real Positivestellensatz, it yields that $P(\theta)\widehat{L}(\theta) + \widehat{L}(\theta)^T P(\theta) > 0$, which ends this proof. □

Note that (8) requires to find appropriate $c_1, c_2, P(\theta)$ and $H(\theta)$ at the same time which is difficult and computationally demanding. Next, we propose an efficient method by using the Square Matrix Representation (SMR). Given a polynomial $f_0(\theta)$ of degree $\deg(f_0)$, $\theta \in \mathbb{R}^r$ and $f_0(\theta) \in \mathcal{P}_{\text{SOS}}$, its SMR is as follows:

$$f_0(\theta) = (*)^T (\bar{F}_0 + C(\delta))\phi(r, d_{f_0}), \tag{10}$$

where $(*)^T AB$ is short for $B^T AB$ given in Section II, $\bar{F}_0$ is denoted by the SMR matrix of $f_0(\theta)$, $r$ is the dimension of $\theta$, $d_{f_0}$ is the smallest integer not less than $\frac{\deg(f_0)}{2}$, i.e., $d_{f_0} = \lceil \frac{\deg(f_0)}{2} \rceil$, $\phi(r, d_{f_0}) \in \mathbb{R}^{l(r, d_{f_0})}$ denotes a power vector which contains all monomials of degree less or equal to $d_{p_0}$, $C(\delta)$ is a parameterization of the space $\mathscr{C} = \{C(\delta) \in \mathbb{R}^{l(r,d_{f_0}) \times l(r,d_{f_0})} : C(\delta) = C^T(\delta), (*)^T C(\delta)\phi(r, d_{f_0}) = 0\}$, in which $\delta \in \mathbb{R}^{\vartheta(r, d_{f_0})}$ is a vector of free parameters. $l(r, d_{f_0})$ and $\vartheta(r, d_{f_0})$ can be obtained similarly as in [26]. For the ease of understanding, an illustration is given:

*Example 1:* Given the polynomial $f_1(\theta) = 7\theta^4 + 2\theta^3 + 4\theta^2 + 6\theta + 9$, we have $d_{f_1} = 2$, $r = 1$ and $\phi(r, d_{f_1}) = (\theta^2, \theta, 1)^T$. Then, $f_1(x)$ can be written as follows:

$$\bar{F}_1 = \begin{pmatrix} 7 & 1 & 0 \\ 1 & 4 & 3 \\ 0 & 3 & 9 \end{pmatrix}, \quad C_1(\delta) = \begin{pmatrix} 0 & 0 & -\delta \\ 0 & 2\delta & 0 \\ -\delta & 0 & 0 \end{pmatrix}. \quad \square$$

This technique can be extended to matrix polynomials. Specifically, let $M(\theta) \in \mathcal{P}_{\text{SOS}}^{s \times s}$ be a symmetric matrix polynomial of size $s \times s$ of degree $\text{Deg}(M)$ in $\theta \in \mathbb{R}^r$ (this means that the highest degree of all the entries of $M(\theta)$ is $\text{Deg}(M)$ in $\theta$), i.e., $\text{Deg}(M) = \max(\deg(M_{ij}))$, $d_M = \lceil \frac{\text{Deg}(M)}{2} \rceil$. Then, $M(\theta)$ can be written as

$$M(\theta) = \Phi(\bar{M} + D(\delta), d_M, s) \tag{11}$$

where $\Phi(\bar{M} + D(\delta), m, s) = (*)^T (\bar{M} + D(\delta))(\phi(r, d_M) \otimes I_s)$, $\bar{M}$ is a symmetric matrix, and $D(\delta)$ is a linear parametrization of the linear subspace $\mathscr{D} = \{D(\delta) \in \mathbb{R}^{l(r,d_M)s \times l(r,d_M)s} : D(\delta) = D^T(\delta), (*)^T D(\delta)(\phi(r, d_M) \otimes I_s) = 0\}$. Now, we can propose the condition of LMIs for checking the connectedness of uncertain networks:

*Theorem 1:* Condition (6) holds if $c^* > 0$, where $c^*$ is the solution of the convex optimization problem

$$c^* = \sup_{c, \bar{R}_i, \bar{P}, \delta} c$$
$$\text{s.t.} \begin{cases} \bar{R}_i \geq 0, \ \bar{P} \geq 0, \ \text{trace}(\bar{P}) = 1 \\ \bar{H} + D(\delta) - cI_{(N-1) \cdot l(r, d_H)} - \sum_{i=1}^{h} \bar{\Upsilon}_i(\bar{R}_i) \geq 0. \end{cases} \tag{12}$$

The matrices involved in this problem are defined by $R_i(\theta) = \Phi(\bar{R}_i, d_{Ri}, N-1)$, $R_i(\theta)s_i(\theta) = \Phi(\bar{\Upsilon}_i(\bar{R}_i), d_{Rs}, N-1)$, $P(\theta) = \Phi(\bar{P}, d_P, N-1)$, $H(\theta) = \Phi(\bar{H} + D(\delta), d_H, N-1)$ in which $2d_{Ri}, 2d_P$ and $2d_H$ are the degrees of $R_i(\theta)$, $P(\theta)$, and $H(\theta) - cI$, respectively.

*Proof:* Suppose that (12) holds. Pre- and post-multiplying the first inequality in (12) by $(\phi(r, d_{Ri}) \otimes I_{N-1})^T$ and $(\phi(r, d_{Ri}) \otimes I_{N-1})$, respectively, one has that

$$R_i(\theta) \geq 0, \forall \theta \in \Omega.$$

Thus, the first constraint in (8) holds by choosing a power vector $\phi(r, d_{Ri}) \otimes I_{N-1}$. Similarly, by pre- and post-multiplying the last inequality in (12) by $(\phi(r, d_H) \otimes I_{N-1})^T$ and $(\phi(r, d_H) \otimes I_{N-1})$, respectively, one has that

$$\begin{aligned} 0 &\leq H(\theta) - c(\phi(r, d_H) \otimes I_{N-1})^T (\phi(r, d_H) \otimes I_{N-1}), \\ &\leq H(\theta) - c((\phi(r, d_H)^T \cdot \phi(r, d_H)) \otimes I_{N-1}), \\ &\leq H(\theta) - (c \cdot \phi(r, d_H)^T \cdot \phi(r, d_H))I_{N-1}, \\ &\leq H(\theta) - \bar{c}_2 I_{N-1}, \end{aligned}$$

where $\bar{c}_2 = (c \cdot \phi(r, d_H)^T \cdot \phi(r, d_H))$. It directly yields that the last constraint in (8) is satisfied. Moreover, let us assume the second and the third constraints in (12) hold. One has that for all the eigenvalues of $\bar{P}$, $\lambda_k(\bar{P}) \geq 0$ and $\sum_{k=1}^{(N-1) \cdot l(r, d_P)} \lambda_k(\bar{P}) = 1$, which results in $\exists k \in \{1, 2, \ldots, (N-1) \cdot l(r, d_P)\}$ such that $\lambda_k(\bar{P}) > 0$. Then, pre- and post-multiplying the second inequality in (12) by $(\phi(r, d_{Ri}) \otimes I_{N-1})^T$ and $(\phi(r, d_{Ri}) \otimes I_{N-1})$, respectively, one has that $P(\theta) > 0$, which means there exists a positive constant $\hat{c} > 0$ and $\bar{c}_1 = (\hat{c} \cdot \phi(r, d_P)^T \cdot \phi(r, d_P))$ such that $P(\theta) - \bar{c}_1 I_{N-1} \geq 0$. Therefore, the condition (8) holds, which completes this proof. □

*B. Multi-Objective Controller Design*

To achieve collision avoidance and connectedness maintenance, Lyapunov-like barrier functions are employed. For the brevity of expressions, let $y_i = x_i - \tau_i$, $y_{ij} = y_i - y_j$, $x_{ij} = x_i - x_j$, $\tau_{ij} = \tau_i - \tau_j$ and $\rho_{ij} = \rho_i - \rho_j$.

For connectedness maintenance, based on Assumption 2, the desired topology is contained in the initial graph. The main idea is to preserve the desired topology $\mathcal{E}^f \subseteq \mathcal{E}(t)$ such that the network is always connected for $t \geq t_0$. To do this, we would like to make the following condition satisfied: $\|x_{ij}\| < r_s$, for all $i \in \mathcal{N}$ and $j \in \mathcal{N}^{\text{sf}}(t)$ which holds if $r_s - \|\tau_{ij}\| - \|y_{ij}\| > 0$. Thus, the following barrier function is used:

$$\Psi_{ij}^{\text{e}} = \frac{\|y_{ij}\|^2}{\hat{r}_s - \|y_{ij}\| + \frac{\hat{r}_s^2}{\mu_1}}, \ i \in \mathcal{N}, \ j \in \mathcal{N}_i^{\text{sf}}(t), \tag{13}$$

where $\hat{r}_s = r_s - \|\tau_{ij}\|$, $\mathcal{N}_i^{\text{sf}}(t) = \{j| j \in \mathcal{N}_i^{\text{s}}(t), j \in \mathcal{N}_i^{\text{f}}\}$ defined in Sectioin II, $\mu_1$ is a positive scalar such that $\Psi_i^{\text{e}}$ is bounded when $\|y_{ij}\|$ tends to $\hat{r}_s$.

For collision avoidance, the basic idea is to keep the distance between any two agents $i$ and $j$ greater than a minimum user-defined safety distance $d_s > 2r_c$. In other words, the condition is required that $\|x_{ij}\| > d_s$, which holds if and only if $\|y_{ij} + \tau_{ij}\| - d_s > 0$. Thus, we select the following barrier function:

$$\Psi_{ij}^{\text{c}} = \frac{(\|y_{ij} + \tau_{ij}\| - \|\tau_{ij}\|)^2}{\|y_{ij} + \tau_{ij}\| - d_s + \frac{(d_s - \|\tau_{ij}\|)^2}{\mu_2}}, \ i \in \mathcal{N}, \ j \in \mathcal{N}_i^{\text{sz}}(t), \tag{14}$$

where $\mathcal{N}_i^{\text{sz}}(t) = \{j|\ j \in \mathcal{N}_i^{\text{s}}(t),\ \|x_{ij}\| < r_z\}$ introduced in Section II. $\mu_1$ is a positive scalar such that $\Psi_i^{\text{e}}$ is bounded when $\|y_{ij}\|$ tends to $d_s$. Therefore, the barrier function considering both collision avoidance and connectedness maintenance can be given as

$$\Psi_{ij} = \frac{\|y_{ij}\|^2}{\hat{r}_s - \|y_{ij}\| + \frac{\hat{r}_s^2}{\mu_1}} + \frac{(\|y_{ij} + \tau_{ij}\| - \|\tau_{ij}\|)^2}{\|y_{ij} + \tau_{ij}\| - d_s + \frac{(d_s - \|\tau_{ij}\|)^2}{\mu_2}}, \quad (15)$$

for $\|x_{ij}\| \in (d_s, r_z)$.

*Remark 2:* We assume $\mu_1$ and $\mu_2$ satisfying $\mu_1 > \mu_{\max}$ and $\mu_2 > \mu_{\max}$ with $\mu_{\max} := \frac{1}{2} \sum_{i=1}^N (\sum_{j \in \mathcal{N}_i^{\text{f}}} \Psi_{ij}^{\text{e}}(\|\hat{r}_s - \hat{\varepsilon}\|) + y_i(t_0)^T \sum_{j=1}^N G_{ij}(t_0) y_{ij}(t_0) + \rho_i(t_0)^T \rho_i(t_0)) + (N-1) N \Psi_{ij}^{\text{c}}(\|d_s - \hat{\varepsilon}\|)$, where $0 < \hat{\varepsilon} < \min\{\frac{1}{2} d_s - r_c, \varepsilon\}$. Barrier function in (15) is different with the existing work in the sense that in [10], [18]–[20], the objective of collision avoidance is not considered; in [20], [23], the control input is not bounded. To make (15) easy to understand, the following example is provided for illustration. □

For the brevity of notations, let us introduce $\Psi_i^{\text{e}} = \sum_{j \in \mathcal{N}_i^{\text{sf}}} \Psi_{ij}^{\text{e}}$, $\Psi_i^{\text{c}} = \sum_{j \in \mathcal{N}_i^{\text{sz}}} \Psi_{ij}^{\text{c}}$, $x = (x_1^T, x_2^T, \ldots, x_N^T)^T$, $\rho = (\rho_1^T, \rho_2^T, \ldots, \rho_N^T)^T$. A distributed parameter-dependent controller is provided as follows:

$$u_i = -\underbrace{\sum_{j \in \mathcal{N}_i^{\text{sf}}(t)} \nabla_{y_i} \Psi_{ij}^{\text{e}}(\|y_{ij}\|)}_{\alpha^{\text{e}}} - \underbrace{\sum_{j \in \mathcal{N}_i^{\text{sz}}(t)} \nabla_{y_i} \Psi_{ij}^{\text{c}}(\|y_{ij}\|)}_{\alpha^{\text{c}}} - \underbrace{\sum_{j \in \mathcal{N}_i^{\text{s}}(t)} G_{ij}(t, \theta) y_{ij}}_{\beta^y} - \underbrace{\sum_{j \in \mathcal{N}_i^{\text{s}}(t)} G_{ij}(t, \theta) \rho_{ij}}_{\beta^\rho}, \quad (16)$$

where $\theta$ is introduced in (3), $G_{ij}$ is the $ij$-th entry of weighted adjacency matrix. The gradient-based terms $\alpha^{\text{e}}$ and $\alpha^{\text{c}}$ are designed to achieve connectedness maintenance and collision avoidance, respectively; the parameter-dependent terms $\beta^y$ and $\beta^\rho$ are designed to achieve the convergence of $y_{ij}$ and $\rho_{ij}$, respectively. The following result shows that under the proposed controller (16), the connectedness of uncertain networks can be ensured.

*Theorem 2:* Under Assumption 1-3, and provided that (12) is satisfied for $\mathcal{G}(t_0, \theta)$, the uncertain graph $\mathcal{G}(t, \theta)$ is connected, and collision avoidance is guaranteed for all $t \geq t_0$ and for all $\theta \in \Omega$ by using the controller (16).

*Proof*: This proof aims to show the concerned set is in a positively invariant set, which implies the connectedness and collision avoidance in the uncertain network. Specifically, we assume that the edge set $\mathcal{E}(t)$ changes at $t_l$, $l = 0, 1, 2, \ldots$. For each $[t_l, t_{l+1})$, $\mathcal{G}$ is fixed. Let us introduce a parameter-dependent Lyapunov-like function

$$W = \frac{1}{2} \sum_{i=1}^N \Bigg( \sum_{j \in \mathcal{N}_i^{\text{sf}}(t)} \Psi_{ij}^{\text{e}}(\|y_{ij}\|) + \sum_{j \in \mathcal{N}_i^{\text{sz}}(t)} \Psi_{ij}^{\text{c}}(\|y_{ij}\|) + y_i \sum_{j=1}^N G_{ij}(t, \theta) y_{ij} + \rho_i^T \rho_i \Bigg). \quad (17)$$

Consider the time interval $[t_0, t_1)$, one has $\Psi_{ij}^{\text{e}} > 0$ from (13), $\Psi_{ij}^{\text{c}}$ from (14), and $\rho_i^T \rho \geq 0$. In addition, $\sum_{i=1}^N y_i \sum_{j=1}^N G_{ij}(t, \theta) y_{ij} = \sum_{i=1}^N y_i \sum_{j=1}^N L_{ij}(t, \theta) y_{ij} = y^T (L(t, \theta) \otimes I_n) y \geq 0$ on account of the fact that $L(t, \theta) = L(t_0, \theta)$, and (12) is satisfied for $\mathcal{G}(t_0, \theta)$. Thus, one has that $W_0 = W(t_0, \theta) > 0$ for all $\theta \in \Omega$. Moreover, for $t \in [t_0, t_1)$, $G_{ij}(t, \theta)$ is fixed, one has

$$\dot{W} = \frac{1}{2} \sum_{i=1}^N \Bigg( \sum_{j \in \mathcal{N}_i^{\text{sf}}(t)} \dot{\Psi}_{ij}^{\text{e}}(\|y_{ij}\|) + \sum_{j \in \mathcal{N}_i^{\text{sz}}(t)} \dot{\Psi}_{ij}^{\text{c}}(\|y_{ij}\|) \Bigg)$$
$$+ \sum_{i=1}^N \dot{y}_i \sum_{j=1}^N L_{ij} y_j + \sum_{i=1}^N \rho_i^T \dot{\rho}_i$$
$$= \sum_{i=1}^N \sum_{j \in \mathcal{N}_i^{\text{sf}}(t)} \dot{y}_i^T \nabla_{y_i} \Psi_{ij}^{\text{e}}(\|y_{ij}\|) + \sum_{i=1}^N \dot{y}_i \sum_{j=1}^N L_{ij} y_j$$
$$+ \sum_{i=1}^N \sum_{j \in \mathcal{N}_i^{\text{sz}}(t)} \dot{y}_i^T \nabla_{y_i} \Psi_{ij}^{\text{c}}(\|y_{ij}\|) + \sum_{i=1}^N \rho_i^T \dot{\rho}_i$$
$$= \sum_{i=1}^N \sum_{j \in \mathcal{N}_i^{\text{sf}}(t)} \rho_i^T \nabla_{y_i} \Psi_{ij}^{\text{e}}(\|y_{ij}\|) + \sum_{i=1}^N \rho_i \sum_{j=1}^N L_{ij} y_j$$
$$+ \sum_{i=1}^N \sum_{j \in \mathcal{N}_i^{\text{sz}}(t)} \rho_i^T \nabla_{y_i} \Psi_{ij}^{\text{c}}(\|y_{ij}\|)$$
$$- \sum_{i=1}^N \rho_i^T \Bigg( \sum_{j \in \mathcal{N}_i^{\text{sf}}(t)} \nabla_{y_i} \Psi_{ij}^{\text{e}}(\|y_{ij}\|)$$
$$+ \sum_{j \in \mathcal{N}_i^{\text{sz}}(t)} \nabla_{y_i} \Psi_{ij}^{\text{c}}(\|y_{ij}\|)$$
$$+ \sum_{j \in \mathcal{N}_i^{\text{s}}(t)} G_{ij}(t, \theta) y_{ij} + \sum_{j \in \mathcal{N}_i^{\text{s}}(t)} G_{ij}(t, \theta) \rho_{ij} \Bigg)$$
$$= -\rho^T (L(t_0, \theta) \otimes I_n) \rho. \quad (18)$$

Taking into account that (12) is satisfied for $\mathcal{G}(t_0, \theta)$, one has $L(t_0, \theta) \geq 0$ for all $\theta \in \Omega$, which implies that $\dot{W} \leq 0$. Thus, $W(t, \theta) \leq W(t_0, \theta) \leq \mu_{\max}$, for $t \in [t_0, t_1)$. From (13), (14) and Remark 2, one has that $\Psi_{ij}^{\text{e}}(\hat{r}_s) = \mu_1 > \mu_{\max}$, and $\Psi_{ij}^{\text{c}}(\hat{d}_s) = \mu_2 > \mu_{\max}$, which implies that no collision occurs during $[t_0, t_1)$, and no agent $j$ has left the set $\mathcal{N}_i^{\text{sf}}$ for agent $i$. Hence, the network $\mathcal{G}(t, \theta)$ is still connected. Let us consider $t = t_1$, we assume that the number of new agents added in the set $\mathcal{N}_i^{\text{sz}}$ is $k_i$ for agent $i$. One has that $\sum_{i=1}^N k_i + \sum_{i=1}^N \text{num}_i(\mathcal{N}_i^{\text{sz}}) \leq N(N-1)$, and $\text{num}_i(\mathcal{N}_i^{\text{sz}})$ is the number of agents in $\mathcal{N}_i^{\text{sz}}$. It yields that

$$W(t_1, \theta) \leq W(t_1^-, \theta) + \sum_{i=1}^{N} k_i \widetilde{\Psi}$$
$$\leq W(t_0, \theta) + \sum_{i=1}^{N} k_i \widetilde{\Psi}$$
$$\leq \frac{1}{2} \sum_{i=1}^{N} \Big( \sum_{j \in \mathcal{N}_i^f} \Psi_{ij}^e(\|\hat{r}_s - \hat{\varepsilon}\|) +$$
$$+ y_i(t_0)^T \sum_{j=1}^{N} G_{ij}(t_0) y_{ij}(t_0) + \rho_i(t_0)^T \rho_i(t_0)$$
$$+ \sum_{j \in \mathcal{N}_i^{sz}(t)} \Psi_{ij}^c(\|y_{ij}\|) \Big) + \sum_{i=1}^{N} k_i \widetilde{\Psi}$$
$$< \mu_{\max}, \tag{19}$$

where $\widetilde{\Psi} = \frac{1}{2} \sum_{j \in \mathcal{N}_i^{sz}} \Psi_{ij}^c(\|\hat{d}_s - \hat{\varepsilon}\|)$.

The above argument can be applied to time intervals $[t_l, t_{l+1})$. The condition still holds that $\dot{W}(t, \theta) \leq 0$, and one has

$$W(t, \theta) \leq W(t_l) \leq \mu_{\max}, \tag{20}$$

which implies that no collision occurs during $[t_l, t_{l+1})$, and no agent $j$ has left the set $\mathcal{N}_i^{sf}$ for agent $i$. Hence, the graph $\mathcal{G}(t, \theta)$ is connected for $t \in [t_l, t_{l+1})$. □

The above result achieves the objectives of last two items in Problem 1. Next, we will show that the objective of the first item in Problem 1 can also be achieved.

*Theorem 3:* If Assumption 1-3 hold and (12) is satisfied for $\mathcal{G}(t_0, \theta)$, then, under the controller (16), the following conditions hold for all $\theta \in \Omega$, $i \in \mathcal{N}$:
  1) $\lim_{t \to \infty} \|\rho_i - \rho_j\| = 0$, for $j \in \mathcal{N}$;
  2) $\lim_{t \to \infty} \|x_i(t) - \tau_i - (x_j(t) - \tau_j)\| = 0$, for $j \in \mathcal{N}_i^f$.

*Proof:* 1) For the first statement, we assume that the edge set $\mathcal{E}(t)$ changes at $t_l$, $l = 0, 1, 2, \ldots$, and there exists a time $\hat{t}_l$ such that the topology of $\mathcal{G}$ is fixed. For $t \in [\hat{t}_l, \infty)$, from the construction of $W$, one has that

$$\frac{1}{2} \sum_{i=1}^{N} y_i \sum_{j=1}^{N} G_{ij}(t, \theta) y_{ij} \leq \mu_{\max}, \quad \frac{1}{2} \sum_{i=1}^{N} \rho_i^T \rho_i \leq \mu_{\max}.$$

Since the topology of $\mathcal{G}$ is fixed, we have that $G_{ij}$ is also fixed for $t \in [\hat{t}_l, \infty)$. On account of the symmetry of $G$, let $\lambda_{\max}$ be the largest eigenvalue of $G$, one has that

$$\frac{1}{2} y^T (L(\hat{t}_l, \theta) \otimes I_n) y \leq \frac{1}{2} \lambda_{\max} \|y\|^2 \leq \mu_{\max},$$

which implies that $\|y\| \leq \sqrt{\frac{2\mu_{\max}}{\lambda_{\max}}}$. Via similar arguments, we obtain that $\|\rho\| \leq \sqrt{2\mu_{\max}}$. Let us consider the set $\Xi = \{y \in \mathbb{R}^{Nn}, \rho \in \mathbb{R}^{Nn} | W(y, \rho) \leq \mu_{\max}, \|y\| \leq \sqrt{\frac{2\mu_{\max}}{\lambda_{\max}}}, \|\rho\| \leq \sqrt{2\mu_{\max}}\}$, which is closed and bounded, and thus a compact set. Now, let us investigate the largest invariant set in $\mathcal{I} = \{y \in \mathbb{R}^{Nn}, \rho \in \mathbb{R}^{Nn} | \dot{W} = 0\}$.

Based on (18), one has

$$\dot{W} = -\rho(L(\theta) \otimes I_n)\rho = -\sum_{i \in \mathcal{N}, j \in \mathcal{N}_i^s} G_{ij} \|\rho_i - \rho_j\|^2,$$

which implies that $\dot{W} = 0$ if and only if $\rho_1 = \cdots = \rho_N$. By using LaSalle's invariance principle, one has that all the trajectories started in the set $\Xi$ will converge to set $\mathcal{I}$, i.e., $\rho_1 = \cdots = \rho_N$.

2) For the second statement, consider the case of $t \geq \hat{t}_l$, one has $\rho_i - \rho_j = 0$ for all $i, j \in \mathcal{N}$. Then, (16) can be rewritten as

$$u_i = -\sum_{j \in \mathcal{N}_i^{sf}(t)} \nabla_{y_i} \Psi_{ij}^e(\|y_{ij}\|) - \sum_{j \in \mathcal{N}_i^{sz}(t)} \nabla_{y_i} \Psi_{ij}^c(\|y_{ij}\|)$$
$$- \sum_{j \in \mathcal{N}_i^s(t)} G_{ij}(t, \theta) y_{ij},$$
$$= -\sum_{j \in \mathcal{N}_i^{sf}(t)} \frac{\partial \Psi_{ij}^e(\|y_{ij}\|)}{\partial \|y_{ij}\|} \cdot \frac{1}{\|y_{ij}\|} y_{ij} - \sum_{j \in \mathcal{N}_i^s(t)} G_{ij}(t, \theta) y_{ij}$$
$$- \sum_{j \in \mathcal{N}_i^{sz}(t)} \frac{\partial \Psi_{ij}^c(\|y_{ij}\|)}{\partial \|y_{ij}\|} \cdot \frac{1}{\|y_{ij}\|} y_{ij}. \tag{21}$$

Observe that $\frac{\partial \Psi_{ij}^e(\|y_{ij}\|)}{\partial \|y_{ij}\|} \cdot \frac{1}{\|y_{ij}\|}$ and $\frac{\partial \Psi_{ij}^c(\|y_{ij}\|)}{\partial \|y_{ij}\|} \cdot \frac{1}{\|y_{ij}\|}$ are positive and bounded as $\|y_{ij}\| \to 0$, one has that $u_i = -(\tilde{L}(t) \otimes I_n + L(\theta, t) \otimes I_n) y$ with $\tilde{L}(t) \geq 0$ and $L(\theta, t) \geq 0$. From algebraic graph theory [3], it yields that $\lim_{t \to \infty} y = \text{span}(1_{Nn})$, i.e., $y_i - y_j = 0$, for all $i, j \in \mathcal{N}$. □

## IV. EXAMPLES

Two examples are provided for illustration. The MATLAB toolbox SeDuMi is used for solving LMI problems.

### A. Example 2: A 6-Agent Case

With an initial topology shown in Fig. 3 (a), an uncertain 6-agent system is considered. The model of agents is set up with the parameters: $r_a = 0.75$, $r_s = 8$, $r_z = 2.5$, $r_c = 1.25 r_a$, $d_s = 2r_c$, and $\epsilon = 0.1$.

It is assumed that the network is affected by two uncertain parameters, i.e. $\theta_1$ and $\theta_2$. Specifically, the uncertain matrix $G(t_0, \theta)$ is given by

$$\begin{bmatrix} 0 & 3+2\theta_1 & 0 & 0 & 0 & 0 \\ 3+2\theta_1 & 0 & 4+\theta_1 & 0 & 0 & 0 \\ 0 & 4+\theta_1 & 0 & 3-\theta_2 & 2\theta_1+\theta_2 & 0 \\ 0 & 0 & 3-\theta_2 & 0 & 0 & 0 \\ 0 & 0 & 2\theta_1+\theta_2 & 0 & 0 & 2 \\ 0 & 0 & 0 & 0 & 2 & 0 \end{bmatrix}.$$

where $\theta \in \mathbb{R}^2$ is constrained in the set $\Omega$ chosen as $\Omega = \{\theta : \|\theta\| \leq 1\}$. Hence, we have $n = 4$ and $r = 2$. For the newly added edge $(i, j)$ for $t > t_0$, we assume $G(i, j) = G(j, i) = 1$. Moreover, $\Omega$ can be described as in (3) with

$$s_1(\theta) = 1 - \theta_1^2 - \theta_2^2.$$

Let us use (6) to investigate the connectedness of network. We look for a constant matrix function $P(\theta)$ satisfying (6), and by solving (12) we find $c^* = 0.589$. Therefore, $\mathcal{G}(t_0, \theta)$ is connected.

Next, we apply the proposed controller (16), and the results are shown in Fig. 4. Let us observe that $\min(x_{ij})$ is always larger than $d_s$ in Fig. 5. Thus, the collision avoidance is achieved. For the connectedness maintenance, by only preserving connections of $j \in \mathcal{N}_i^{\text{sf}}$, the controller allows breaks of edges when system evolves. Demonstrated by Fig. 3-4, our approach guarantees the connectedness of the uncertain network, and the robust formation is achieved.

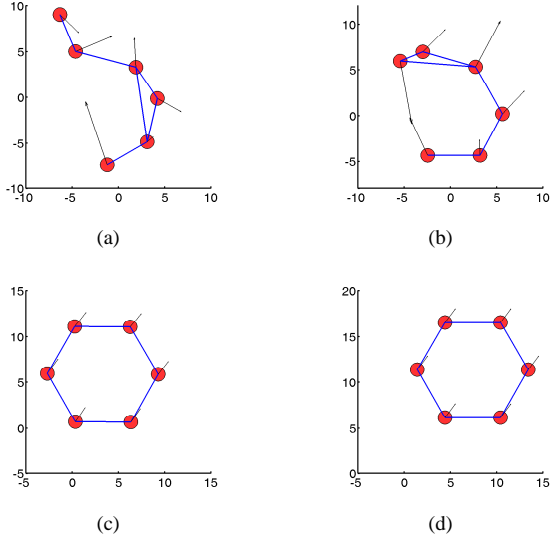

Fig. 3. Example 2: The motion of agents and the set of edges of network for $t = 0, 1, 3, 4$, respectively.

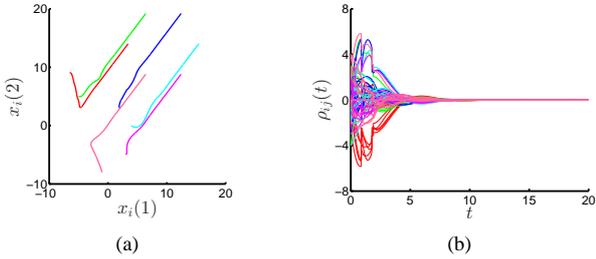

Fig. 4. Example 2: The trajectories of agents and the differences of velocities.

### B. Example 3: A 50-Agent Case

In this scenario, a 50-agent system is considered with the parameters: $N = 50$, $r_a = 0.75$, $r_s = 5$, $r_z = 2.5$, $r_c = 1.25 r_a$, $d_s = 2r_c$, and $\epsilon = 0.1$. The initial positions are selected as shown in Fig. 6, and the initial velocities are randomly chosen in $[-2, 2]$. The desired configuration is a circle of radius 30 on which all agents are symmetrically distributed. The communication are affected by three parameters. Due to limited space, we omit the expression of $\Omega$ and $G(\theta)$. By solving (12) we find $c^* = 0.642$, which ensures that $\mathcal{G}(t_0, \theta)$ is connected. The proposed controller (16) guarantees the collision avoidance, and the robust formation is achieved as shown in Fig. 7-8. Finally, it is worth noting that the proposed method is scalable for a large number of agents.

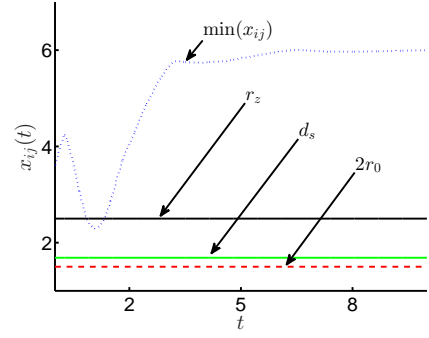

Fig. 5. Example 2: The minimal distance between agents.

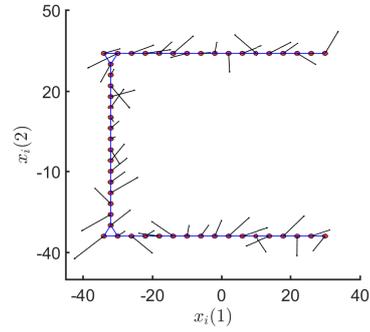

Fig. 6. Example 3: The initial topology and states.

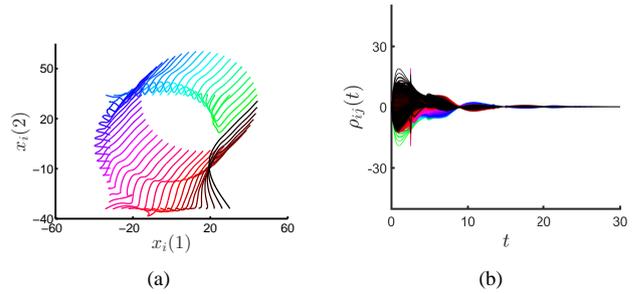

Fig. 7. Example 3: The trajectories of agents and the differences of velocities.

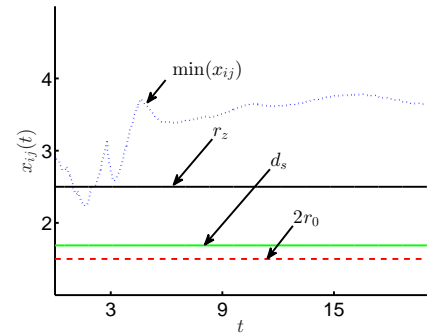

Fig. 8. Example 3: The minimal distance between agents.

## V. Conclusion and Discussion

This paper addresses the robust formation control problem of multiple agents under parametric communication uncertainties. A distributed controller is proposed by introducing a parameter-dependent Lyapunov-like barrier function, which encodes control objectives like connectedness maintenance, collision avoidance and convergence to the desired configuration. Firstly, a necessary and sufficient condition is proposed for checking the connectedness of uncertain networks under communication uncertainties. Based on this condition, a solvable condition consisting of LMIs is given by employing the square matrix representation. Secondly, a parameter-dependent Lyapunov-like barrier function is introduced, and a gradient-based controller with bounded input is designed such that the collision avoidance and connectedness maintenance can be achieved. Finally, this paper also shows that the systems converges to the desired formation configuration by using LaSalle invariance principle.

For the connectedness maintenance, this paper assumes that the edge set of desired formation configuration is contained in the initial graph. This strategy allows the breaks of edges for agent not belonging to set $\mathcal{N}_i^{\text{sf}}$, and provides an efficient way for barrier function construction. We would like to relax this assumption and consider a more general switching strategy used in flocking problem proposed in [25]. In addition, more efforts will be devoted to the comparisons to analytic vector field methods [27], [28], and contraction theory methods [29].

## References

[1] R. Olfati-Saber, J. A. Fax, and R. M. Murray, "Consensus and cooperation in networked multi-agent systems," *Proceedings of the IEEE*, vol. 95, no. 1, pp. 215–233, 2007.

[2] W. Ren and R. Beard, "Consensus seeking in multiagent systems under dynamically changing interaction topologies," *IEEE Transactions Automatic Control*, vol. 50, no. 5, pp. 655–661, 2005.

[3] M. Mesbahi and M. Egerstedt, *Graph theoretic methods in multiagent networks.* Princeton University Press, 2010.

[4] H. G. Tanner, A. Jadbabaie, and G. J. Pappas, "Flocking in fixed and switching networks," *IEEE Transactions on Automatic control*, vol. 52, no. 5, pp. 863–868, 2007.

[5] E. G. Hernández-Martínez and E. Aranda-Bricaire, *Convergence and collision avoidance in formation control: A survey of the artificial potential functions approach.* INTECH Open Access Publisher, 2011.

[6] K. Oh, M. Park, and H. Ahn, "A survey of multi-agent formation control," *Automatica*, vol. 53, pp. 424–440, 2015.

[7] M. Cao, C. Yu, and B. Anderson, "Formation control using range-only measurements," *Automatica*, vol. 47, no. 4, pp. 776–781, 2011.

[8] D. V. Dimarogonas and K. J. Kyriakopoulos, "A connection between formation infeasibility and velocity alignment in kinematic multi-agent systems," *Automatica*, vol. 44, no. 10, pp. 2648–2654, 2008.

[9] ——, "On the rendezvous problem for multiple nonholonomic agents," *IEEE Transactions on Automatic Control*, vol. 52, no. 5, pp. 916–922, 2007.

[10] H. Su, X. Wang, and G. Chen, "Rendezvous of multiple mobile agents with preserved network connectivity," *Systems & Control Letters*, vol. 59, no. 5, pp. 313–322, 2010.

[11] L. Cheng, Z. Hou, and M. Tan, "A mean square consensus protocol for linear multi-agent systems with communication noises and fixed topologies," *IEEE Transactions on Automatic Control*, vol. 59, no. 1, pp. 261–267, 2014.

[12] Z. Meng, Z. Lin, and W. Ren, "Robust cooperative tracking for multiple non-identical second-order nonlinear systems," *Automatica*, vol. 49, no. 8, pp. 2363–2372, 2013.

[13] G. Hu, "Robust consensus tracking of a class of second-order multi-agent dynamic systems," *Systems & Control Letters*, vol. 61, no. 1, pp. 134–142, 2012.

[14] H. L. Trentelman, K. Takaba, and N. Monshizadeh, "Robust synchronization of uncertain linear multi-agent systems," *IEEE Transactions on Automatic Control*, vol. 58, no. 6, pp. 1511–1523, 2013.

[15] M. Huang and J. H. Manton, "Coordination and consensus of networked agents with noisy measurements: Stochastic algorithms and asymptotic behavior," *SIAM Journal on Control and Optimization*, vol. 48, no. 1, pp. 134–161, 2009.

[16] Z. Li, Z. Duan, and F. L. Lewis, "Distributed robust consensus control of multi-agent systems with heterogeneous matching uncertainties," *Automatica*, vol. 50, no. 3, pp. 883–889, 2014.

[17] D. Panagou, D. M. Stipanović, and P. Voulgaris, "Distributed coordination control for multi-robot networks using Lyapunov-like barrier functions," *IEEE Transactions on Automatic Control*, vol. 61, no. 3, pp. 617–632, 2016.

[18] Y. Dong and J. Huang, "A leader-following rendezvous problem of double integrator multi-agent systems," *Automatica*, vol. 49, no. 5, pp. 1386–1391, 2013.

[19] Z. Feng, C. Sun, and G. Hu, "Robust connectivity preserving rendezvous of multi-robot systems under unknown dynamics and disturbances," *IEEE Transactions on Control of Network Systems*, 2016.

[20] M. Ji and M. Egerstedt, "Distributed coordination control of multiagent systems while preserving connectedness," *IEEE Transactions on Robotics*, vol. 23, no. 4, pp. 693–703, 2007.

[21] D. V. Dimarogonas and K. J. Kyriakopoulos, "Connectedness preserving distributed swarm aggregation for multiple kinematic robots," *IEEE Transactions on Robotics*, vol. 24, no. 5, pp. 1213–1223, 2008.

[22] Z. Kan, A. P. Dani, J. M. Shea, and W. E. Dixon, "Network connectivity preserving formation stabilization and obstacle avoidance via a decentralized controller," *IEEE Transactions on Automatic Control*, vol. 57, no. 7, pp. 1827–1832, 2012.

[23] M. M. Zavlanos, A. Jadbabaie, and G. J. Pappas, "Flocking while preserving network connectivity," in *Proceedings of the Conference on Decision and Control*, 2007, pp. 2919–2924.

[24] M. M. Zavlanos and G. J. Pappas, "Distributed connectivity control of mobile networks," *IEEE Transactions on Robotics*, vol. 24, no. 6, pp. 1416–1428, 2008.

[25] M. M. Zavlanos, M. B. Egerstedt, and G. J. Pappas, "Graph-theoretic connectivity control of mobile robot networks," *Proceedings of the IEEE*, vol. 99, no. 9, pp. 1525–1540, 2011.

[26] G. Chesi, "LMI techniques for optimization over polynomials in control: A survey," *IEEE Transactions on Automatic Control*, vol. 55, no. 11, pp. 2500–2510, 2010.

[27] D. Panagou and V. Kumar, "Cooperative visibility maintenance for leader–follower formations in obstacle environments," *IEEE Transactions on Robotics*, vol. 30, no. 4, pp. 831–844, 2014.

[28] D. Panagou, H. G. Tanner, and K. J. Kyriakopoulos, "Control of nonholonomic systems using reference vector fields," in *IEEE Conference on Decision and Control and European Control Conference*, 2011, pp. 2831–2836.

[29] D. Han and G. Chesi, "Robust synchronization via homogeneous parameter-dependent polynomial contraction matrix," *IEEE Transactions on Circuits and Systems I: Regular Papers*, vol. 61, no. 10, pp. 2931–2940, 2014.